\documentclass[doublecol]{epl2} 
\usepackage{geometry} 
\usepackage{graphicx,color}
\usepackage{graphicx}
\usepackage{amssymb,amsmath}
\usepackage{float}
\usepackage{longtable}
\usepackage{braket}
\usepackage{comment}
\usepackage[mathcal]{euscript}
\title{Origami launcher}
\shorttitle{Origami launcher}
\author{Oryna Ivashtenko\inst{1}, Polina Kofman\inst{1}, Oleksiy Golubov\inst{1,2}, Zakhar Maizelis\inst{1,3}}
\institute{                    
\inst{1} V. N. Karazin Kharkiv National University -- Svobody Sq. 4, 61022, Kharkiv, Ukraine\\    
\inst{2} Department of Aerospace Engineering Sciences, University of Colorado at Boulder, 429 UCB, Boulder, CO, 80309, USA\\
\inst{3} A. Ya. Usikov Institute for Radiophysics and Electronics,
National Academy of Sciences of Ukraine, 12 Proskura Str., Kharkiv, 61085, Ukraine
}

\pacs{nn.mm.xx}{First pacs description}

\abstract{The article studies the elastic and locomotive properties of Miura-ori-type paper origami.
The mechanics of a single paper crease is studied experimentally, and its non-elastic properties turn out to be crucial.
The entire origami construction is then described as a collection of individual creases, its capability to launch small objects is evaluated, and the equation of motion is found. Thus, the height of the launched ball is studied theoretically and experimentally as a function of governing parameters.}

\begin{document}
	\maketitle
	\label{firstpage}

	\section{Introduction}

In our investigation, we will consider the properties of paper origami, designed to vertically launch small loads. This problem was proposed by the organizing committee of the X International Physicists' Tournament. In its original version, we were to optimize a paper construction such as Miura-ori to vertically launch a standard Ping-Pong ball to a maximally possible height. We were also restricted to use only one uncut sheet of A4 paper (80 g/m$^2$). It had a fixed thickness, density and inherent anisotropy caused by the fibrous structure. 

\begin{figure}
\center{\includegraphics[width=1\linewidth]{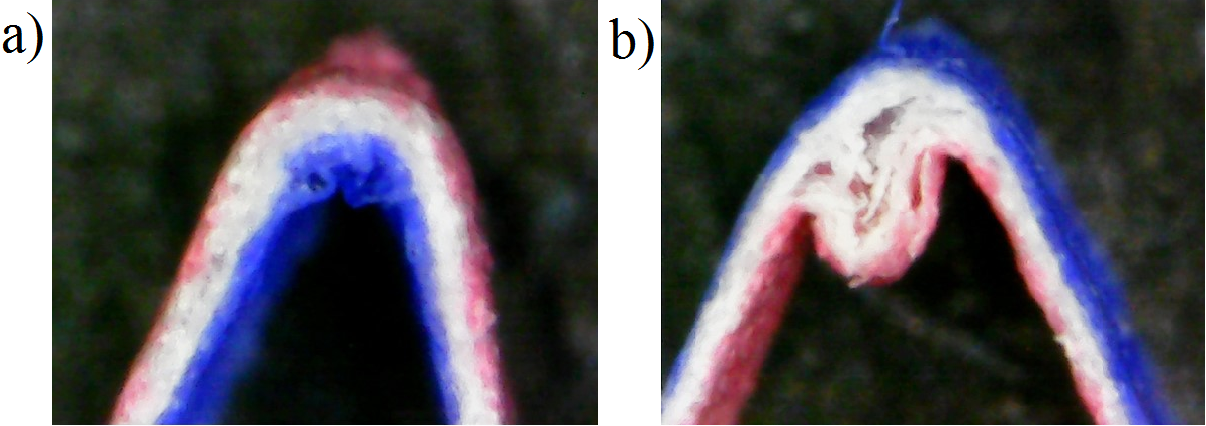}}
\vspace*{-0.5 cm}
\caption{Demonstration of modifications in the structure of the folded paper. a) The first folding of an initially flat sheet of paper. One can see the corrugation of the internal layer. b) The same sample bent backwards. Now the preliminary stretched layer forms many more creases.}
\vspace*{-0.5 cm}
\label{fig.structure}
\end{figure}

We define origami as a paper construction created by folding paper. The creation of creases is accompanied by damage to the material and plastic deformations. This process is irreversible, as can be seen from the photo (fig.~\ref{fig.structure}). However, once folded, origami may be subjected to further deformations that can be elastic in the vicinity of the stable position. Different types of these deformations are possible: bending, stretching, all-round compression etc. Although both stretching and all-round compression can store a lot of energy, we will focus on the optimization of a one-layer Miura-ori, which operates solely on bending deformations.

	\subsection{Design of the folding} 
    Miura-ori is one of the popular models known from the mechanical modeling of metamaterials \cite{Silverberg2014}, \cite{schenk2013}, \cite{schenk2010}.  The Miura fold is a rigid origami, i.e. each cell is flat and does not change either its shape or its linear dimensions during any transformations of the folding. That is why in our treatment we will consider only bending deformations in the edges and require that the energy stored in bending is much greater than the energy stored in other types of deformations. This principle also allows us to treat the construction as a sum of single folds. We note that multilayer structures will be disregarded because of stretching deformations appearing in them.
    
\begin{figure}
\center{\includegraphics[width=1\linewidth]{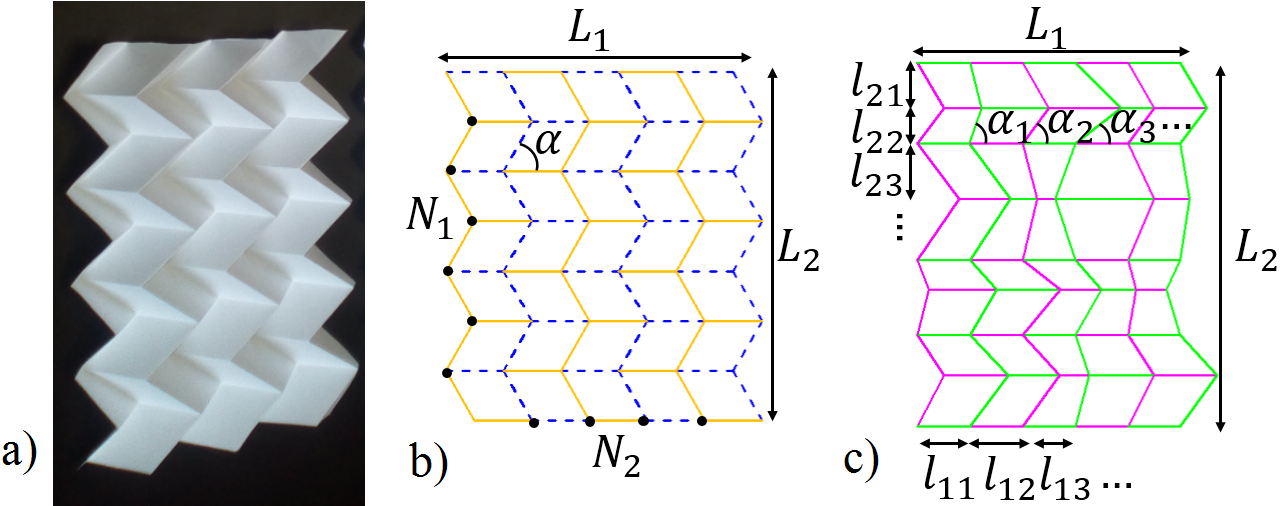}}
\vspace*{-0.5 cm}
\caption{The Miura pattern and parametrization a) Photo of the standard (flat) Miura origami. b) The standard Miura scheme and its parameters: the length $L_1$ and the height $L_2$ of the sheet (fixed), the number $N_1$ of horizontal and $N_2$ of vertical creases, and the angle of construction $\alpha$ (variable). c) The Miura-like structure scheme. All cells may be arbitrarily varied, but $L_1$ and $L_2$ are still fixed.}
\label{fig.schemes}
\end{figure}

    Standard Miura-ori is a pattern consisting of equal parallelograms connected by folds (fig.~\ref{fig.schemes},~a,~b). Besides a standard flat Miura-ori (fig.~\ref{fig.schemes},~a) there are multiple isomorphic variations \cite{Gilewski2014}, \cite{Sareh2015}, leading to a wide range of so called Miura-like structures. We call the standard Miura pattern flat, because Miura-like structures in a folded state may have different shapes, e.g. cylindrical. To construct a mathematical model of the Miura-ori folding pattern we parametrized it using 5 parameters: the length $L_1$ and the height $L_2$ of the paper sheet (which are fixed for our problem), the number $N_1$ of horizontal folds, the number $N_2$ of vertical folds, and the angle of construction $\alpha$ (which are variable). This set of parameters can describe the standard Miura pattern (fig.~\ref{fig.schemes},b). Moreover, by varying angles and distances, one can construct many possible Miura-like structures (fig.~\ref{fig.schemes},c) that can be non-homogeneous or even non-periodic.  
    
    \section{Dead ends}
In this section we try to evaluate the effectiveness of the origami launcher purely theoretically, by applying the linear theory of elasticity and avoiding the use of any phenomenological laws and experimentally determined dependencies. Due to the peculiarities of the paper structure and the presence of very large deformations, this theory could provide at most an estimate of the height reached by the ball.

From the linear theory of beams, we can write the expression for the elastic potential energy of one fold \cite{Timoshenko}:
	\begin{equation} 
		W = \frac{{\phi}^2 IE}{2l}\,.
		\label{W_Timoshenko}
	\end{equation}
Here $\phi=\pi$ is the bending angle of the fold. $E$ is the Young's modulus of the paper. $l=\pi\frac{d}{2}$ denotes the length of the neutral line, with $d$ being the paper thickness. $I=\frac{a d^3}{12}$ is the geometric moment of inertia of the paper's cross-section, with $a=L N_1+ \frac{L_2 N_2}{\sin \alpha}$ being the total length of all the creases on the sheet of paper (fig.~\ref{fig.schemes}). Substituting all these expressions into (\ref{W_Timoshenko}), we obtain:
	\begin{equation} 
		W = \left(L N_1+ \frac{L_2 N_2}{\sin \alpha}\right) d^2 \frac{E \pi}{12}\,.
		\label{Epot}
	\end{equation}

The height to which the ball rises can be estimated from the law of conservation of energy, by equating (\ref{Epot}) to the kinetic energy of the system (\ref{Ek}), thus obtaining the initial velocity of the ball $v$, and expressing the maximum height using the law of uniformly accelerated motion, $h_\mathrm{max}=\frac{v^2}{2g}$. The answer is
	\begin{equation} 
		h_\mathrm{max} = \frac{(L_1N_1\sin{\alpha}+L_2N_2)d^2E \pi}{4\sin{\alpha}(3m+m_0)g}\,.
	\end{equation}

 This expression correctly reproduces some features seen experimentally in fig. \ref{exp}: linear dependence of $h_\mathrm{max}$ on $N_1$ and $N_2$, monotonous decrease of $h_\mathrm{max}$ as a function of $\alpha$.
 Still, assuming $E\approx 3\cdot 10^9$ Pa \cite{Alan1967} and substituting the typical geometric properties from our experiments, we get the value $h_\mathrm{max}\approx 1$ km, which is 3 orders of magnitude too high. The reason for this discrepancy can be seen in fig. \ref{fig.structure}: the inner side of the fold does not compress but corrugates, which also causes a much smaller stretching of the outer side of the fold, with an $E_\mathrm{p}$ much smaller than (\ref{Epot}). Moreover, the real curvature radius of the fold is much larger than the assumed value of $\frac{d}{2}$. To some extent, these deficiencies of the theory can be remedied by introducing into (\ref{W_Timoshenko})-(\ref{Epot}) an ``effective thickness" much smaller than $h$ and an ``effective length" much larger than $l=\pi\frac{d}{2}$. The unavoidable need for such arbitrary fitting parameters and a scandalously large height reached by the ball, in our opinion, render the existence of a simple axiomatic theory highly implausible. We thus consider this unsuccessful attempt to be a motivation for the creation of the phenomenological theory, as presented in the following part of this article.
 
    \section{Methods}
    \subsection{Dynamics of a single paper crease}
     The properties of the paper are related to numerous elastic and frictional processes in its fibrous micro-structure. The torque provided by the bent paper nonlinearly depends on the angle between the facets. Moreover there is a hysteresis (from the cycle of opening/closing the fold), so the properties of paper are process-depending. Moving layers are subjected to the internal friction that also depends on the motion rate. The physics behind all these values is very complicated and sensitive to the particular conditions of the experiment, such as the brand and the quality of the paper, the state of the environment, the folding method etc. That is why it is very difficult to study from microscopics the elastic properties of such a complex material as paper. However, it is possible to construct a simple phenomenological theory describing this system.
     
     To do this, we decided firstly to study the dynamics of a single paper crease, to obtain its basic elastic characteristics from the experiment, and then to apply this result to describe the entire origami. 
    Let us write the equation of motion for the system consisting of a single paper crease fixed by one face to the floor, with an additional load fixed to the edge of its moving face:
    	\begin{equation} 
		I\ddot{\phi} = M(\phi, \dot{\phi})-M_\mathrm{g}(\phi),
        \label{eq_mot_nod}
	\end{equation}
where $I$ is the moment of inertia of the moving part of the paper and the load, $M(\phi, \dot{\phi})$ is the effective torque due to the elastic and frictional forces in the crease, and $M_\mathrm{g}$ is the torque related to the gravity force acting on the load and the paper. In our approach, we assume that the effective torque in its turn consists of the static torque $M_0$ and the dissipative term that depends on the angle between the two paper faces and the angular velocity,
	\begin{equation} 
		M(\phi, \dot{\phi}) = M_0(\phi)+\mu(\phi)\dot{\phi}.
        \label{M}
	\end{equation}
    Here we assume that $M_0$ is only a function of the angle $\phi$, and that the dissipative term is proportional to the angular velocity $\dot{\phi}$ (some analogy to a viscoelastic model) with the coefficient of proportionality $\mu$ depending in its turn on the angle $\phi$. Substituting these expressions and treating the values of the moment of inertia and the gravitational term through the paper and the load parameters, we obtain the differential equation of motion:    
    \begin{align}
    \label{eq_mot_one}
		&\left(ml^2+m_0\frac{L^2}{3}\right)\ddot{\phi}=\\
        &=M_0(\phi)-\mu(\phi)\dot{\phi}-\left(ml+m_0\frac{L}{2}\right)g\cos \phi,\nonumber
	\end{align}   
where $m$ and $l$ are respectively the mass and the lever arm of the load, while $m_0$ and $L$ are the mass and the length of the moving part of the paper.     
    In this equation, the static torque $M_0(\phi)$ and the dynamical dissipative coefficient $\mu(\phi)$ as functions of the angle are to be determined from the experimental study . 
 
 \begin{figure}
 \centering{\includegraphics[width=0.8\linewidth]{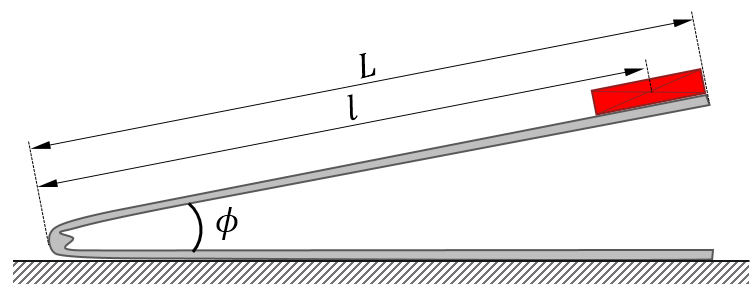}}
 \vspace*{-0.2 cm}
\caption{Scheme of the experimental technique with a single fold. The load (red rectangle) is put at the extremity of one face, while the other is clamped to the surface. Considering the faces to be rigid (indeed, the crease is much easier to bend that the undamaged paper, as we saw in the experiment), we track the temporal evolution $\phi(t)$ of the crease opening, starting from $\phi=0$.}
\vspace*{-0.3 cm}
\label{experiment}
\end{figure}
    
\begin{figure}
 \centering{\includegraphics[width=0.9\linewidth]{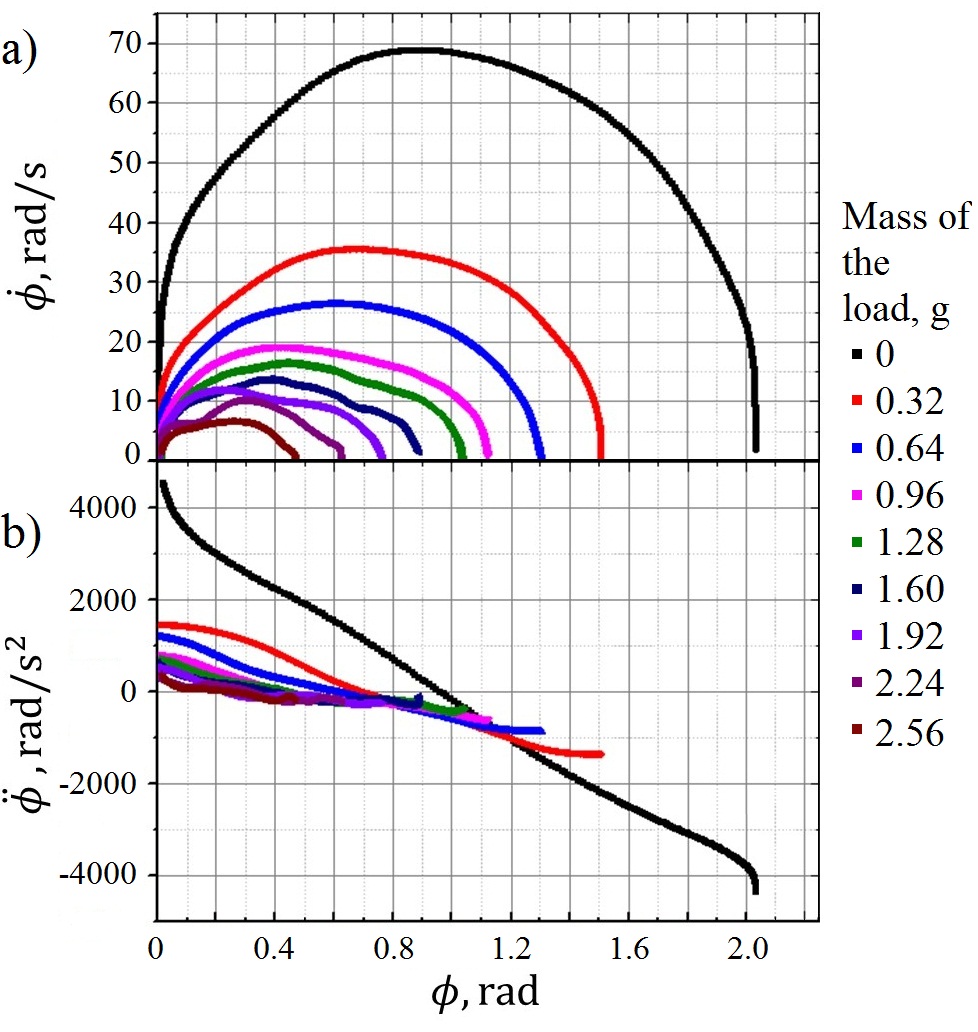}} 
\vspace*{-0.5cm}
\caption{Angular dependences of a) the angular velocity and b) the acceleration. We used the measured time-dependences of the angle, the angular velocity and the angular acceleration to express the latter two as functions of the angle. Different curves correspond to different masses of the load.}
\vspace*{-0.3 cm}
\label{phidot_phiddot}
\end{figure}

    In our experiment (fig.~\ref{experiment}), we filmed the dynamics of a loaded paper crease, traced it using the program \textit{Tracker} \cite{tracker}, thus obtaining the time dependence of the angle $\phi(t)$ for different masses of the load, numerically differentiated them and treated the angular velocity $\dot{\phi}(t)$ and acceleration $\ddot{\phi}(t)$ as functions of $\phi$ (fig.~\ref{phidot_phiddot}). Before performing the numerical differentiation, the experimental data were subjected to smoothing, using spline interpolation. All the points in our experiment come from a single paper sample. These dependencies are reproducible, but they are very sensitive to any changes: paper should be taken from the same package, fold should be made in the same way by the same person (because different pressure during folding etc. causes different damage to the material), environmental conditions such as humidity should also be the same etc. Therefore, unfortunately, to repeat all the following computations in different circumstances one needs to obtain corresponding characteristics of a single bend once again.

\begin{figure}
 \centering{\includegraphics[width=1\linewidth]{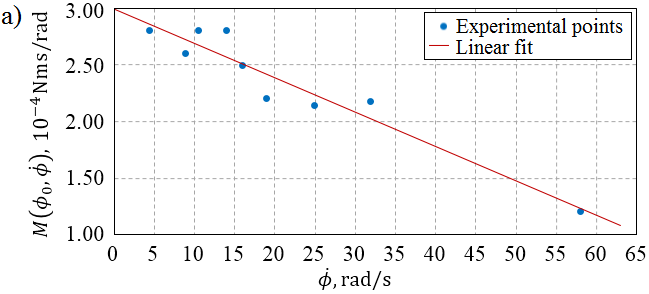}} \\
 \vspace*{0.3cm}
 \centering{\includegraphics[width=1\linewidth]{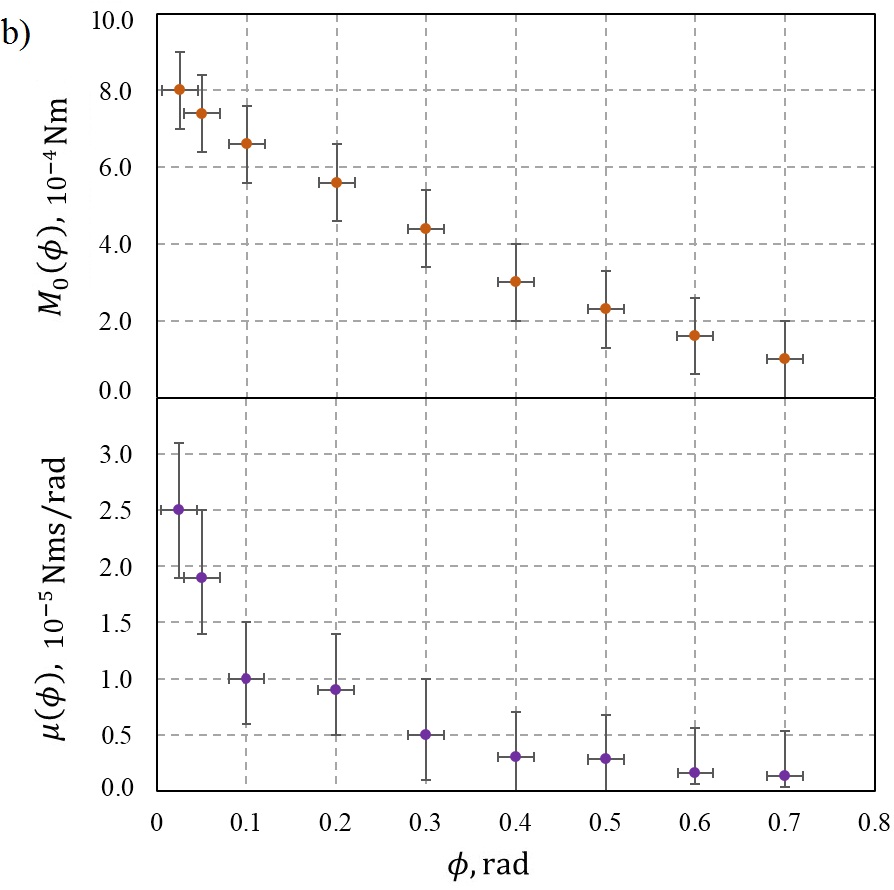}} \\
\vspace*{-0.3cm}
\caption{Processing the experimental data. a) Example of the dependence of the effective torque $M(\phi_0, \dot{\phi})$ on the angular velocity $\dot{\phi}$ for the fixed opening angle $\phi_0\!=\!0.4\,\mathrm{radians}$. Variation of the effective torque at a fixed angle is achieved by changing the load mass. This dependence is fitted according to our assumption $M(\phi,\dot{\phi})\!=\!M_0(\phi)+\mu(\phi)\dot{\phi}$ with $M_0(\phi)$ and $\mu(\phi)$ as fitting parameters. So, from this linear fit we get $M_0(\phi_0)$ and $\mu(\phi_0)$. The same procedure is then repeated for other values of $\phi_0$. 
b) The resulting empirical parameters of our model as functions of the angle of inclination of the paper. The upper plot is the static torque provided by the paper crease and the lower one is the dynamical dissipative parameter slowing down the motion of the paper. The latter is the coefficient of proportionality relating the effective torque and the angular velocity. It should be noted that in further calculations these values will be normalized according to the corresponding length of the crease.}
\label{fig.disscoeff}
\end{figure}

    We analyzed the obtained experimental data to find the empirical parameters of our model. For this purpose, we substituted $\dot{\phi}(\phi)$ and $\ddot{\phi}(\phi)$ into equation (\ref{eq_mot_one}) and treated the effective torque in each case. When varying the mass of the load, the angular velocities are different for the same value of the angle. As such, we fixed the value of the angle, constructed the dependence of the effective torque on the angular velocity, and fitted these data using our linear assumption with $M_0$ and $\mu$ as fitting parameters (fig.~\ref{fig.disscoeff},~a)). The linearity of this dependence verifies our assumption regarding the viscoelastic form of the effective torque of $M(\phi, \dot{\phi})$.

    Repeating the same procedure for different fixed angles we received the static torque $M_0(\phi)$ and the dynamical dissipative coefficient $\mu(\phi)$ as functions of the angle (fig.~\ref{fig.disscoeff},~b)). After interpolation, these functions can be used to solve the equations of motion of complex origami structures consisting of a number of creases.       
    
    \subsection{Operation of the construction as a whole}
    Relying on the obtained results for a single paper crease, we now can build a theoretical model for the operation of a complex origami structure. We treat the origami folding as a set of single creases, so we need only to sum up the work of all these creases, taking into consideration the construction properties. Let us consider the energy transformations of the system including the origami and the load.      
    
     \begin{align} 
		\sum_{i, \, edges} M(\phi_i, \dot{\phi_i})\Delta\phi_i=\Delta E_\mathrm{p}+\Delta E_\mathrm{k}+\Delta E_\mathrm{l},
        \label{En_trans}
	\end{align}   
where $\Delta E_\mathrm{p}$ and $\Delta E_\mathrm{k}$ are respectively the increments of the potential and kinetic energy of both the paper construction and the load, and $\Delta E_\mathrm{l}$ describes all other energy losses related to the load, e.g. air drag, rotation etc. By a simple estimation (see Appendix 1) it can be shown that these losses are negligible with respect to the other terms of the equation, so we will not take $\Delta E_\mathrm{l}$ into consideration in our further computations.

\begin{figure}
\center{\includegraphics[width=1\linewidth]{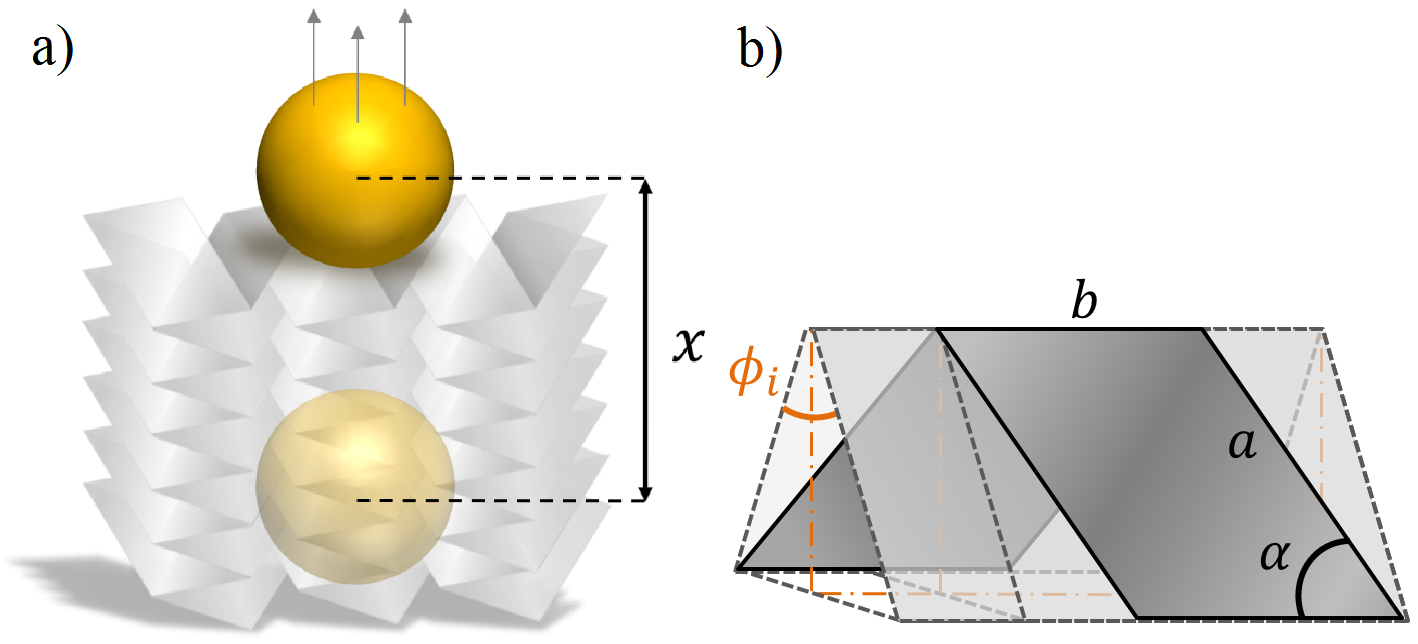}}
\vspace*{-0.6 cm}
\caption{Miura-ori launching a ball. a) The construction disposed vertically with a ball on it. $x$ is the vertical coordinate of the ball measured from the initial position of its center of mass, the origami being in the folded state. b) A single crease within the entire construction. Each of the angles $\phi_i$ corresponds to one of the types of folds. For the case of a standard flat Miura-ori presented in this figure, there are two different types of folds: near the side $a$ and near the side $b$.}
\vspace*{-0.3 cm}
\label{bal_ang}
\end{figure}

Our task is to study the height of the load, so it would be reasonable to rewrite all of the equations in terms of the vertical coordinate $x$ (fig~\ref{bal_ang},~a). To do this, we need to establish relations between the change of the coordinate $x$ and that of each of the angles $\phi_i$. The elementary work of the effective torque then reads

 \begin{align} 
		\mathcal{M}(x,\dot{x}) dx=\hspace{-0.6em}\sum_{i, \,edges}\hspace{-0.5em} M_0(\phi_i(x))\frac{d\phi_i}{dx} dx-\nonumber\\
        -\hspace{-0.5em}\sum_{i,  \,edges}\hspace{-0.5em} \mu(\phi_i(x))\dot{x}\left(\frac{d\phi_i}{dx}\right)^2\!dx  \,,
        \label{M_nod(x)}
 \end{align}
where $M_0$ and $\mu$ are numerical functions (fig.~\ref{fig.disscoeff},~b)) normalized according to the length of the creases.

To find the coefficients of transformation $\frac{d\phi_i}{dx}$, we need to express the angles $\phi_i$ in terms of the coordinate $x$. Firstly, we express the angles between the two particular facets in terms of the local displacement of one fold, and then we generalize this result to the entire construction with the total displacement $x$. In the case of a standard flat Miura-ori, we have two different types of angles (fig.~\ref{bal_ang},~b). We solve this geometric problem (see Appendix 2) and obtain the following expressions:
 \begin{equation} 
		\phi_1 = \arcsin\left[2 \frac{x}{L_2}\; \sqrt[]{1-\frac{x^2}{L_2^2}}\right],
        \label{phi1}
 \end{equation}
  \begin{equation} 
        \phi_2 = \arcsin\left[\frac{2 \cos\alpha\ \sqrt[]{\frac{L_2^2}{x^2}-1}}{\frac{L_2^2}{x^2}-\sin^2\alpha}\right].
        \label{phi2}
 \end{equation}
As can be seen, these expressions are related to the parameters of the particular pattern discussed in the previous section. Here, $L_2$ is the height of the paper sheet and $\alpha$ is the construction angle (see fig.~\ref{fig.schemes},~b). For the standard flat Miura pattern we have only one type of cell, and hence only one construction angle. We also need to find the derivatives of the obtained relations. Differentiating (\ref{phi1}) and (\ref{phi2}) we get
 \begin{equation} 
		\frac{d\phi_1}{dx} = \frac{2}{\sqrt[]{L_2^2-x^2}},
        \label{phidot1}
 \end{equation}
 \vspace*{-0.1 cm}
 \begin{equation} 
        \frac{d\phi_2}{dx} = \frac{2 \cos\alpha\ }{\sqrt[]{L_2^2-x^2}\left(1-\frac{x^2}{L_2^2}\sin^2\alpha\right)}.
        \label{phidot2}
 \end{equation}
Using these expressions (\ref{phi1}-\ref{phidot2}), $\mathcal{M}(x,\dot{x})$ can be treated in terms of $x$: 
\vspace*{-0.2 cm}
 \begin{align}
 \label{M(x)}
		&\mathcal{M}(x,\dot{x})=\\
        &=N_1L_1\Bigg(\!M_0[\phi_1(x)]\frac{d\phi_1}{dx}\!-\mu[\phi_1(x)]\dot{x}\!\left(\!\frac{d\phi_1}{dx}\!\right)^2\!\Bigg)+\nonumber\\
        &+\frac{N_2L_2}{\sin \alpha}\Bigg(\!M_0[\phi_2(x)]\frac{d\phi_2}{dx}\!-\mu[\phi_2(x)]\dot{x}\!\left(\!\frac{d\phi_2}{dx}\!\right)^2\!\Bigg),\nonumber
 \end{align}
where $N_1$ and $N_2$ are respectively the numbers of horizontal and vertical folds. The multipliers $L_1$ and $\frac{N_2L_2}{\sin \alpha}$ come from calculating the length of the folds of a given type (see fig.~\ref{fig.schemes},~b). 

In respect of the kinetic and potential energy terms $E_\mathrm{k}$ and $E_\mathrm{p}$, they will consist of the energies of the load and of the origami itself. The expression for the total kinetic energy of the origami is complicated as it includes all of the inner motions of each particular crease. To simplify the calculations, we will neglect these inner motions and consider only the vertical velocity of the paper. Of course, this neglect may result in errors of our computations, but we accept it as the first approximation. Thus, the kinetic energy of the origami also can be easily treated in terms of the vertical motion.

We assume that the bottom of the origami remains at rest (we fix it to the table), its top moves at the speed of the ball $v$, and the vertical velocity of the paper changes linearly with height (in reality, it is not exactly so because of the increasing effective load on the lower folds due to the mass of the above paper). Then, averaging the squared velocity over the length of the origami, we get the following expression for the kinetic energy of the system:
  \begin{equation}
  E_\mathrm{k}=E_{\mathrm{k,ball}}+E_{\mathrm{k,origami}}=\frac{m v^2}{2}+\frac{1}{3} \frac{m_0 v^2}{2},
  \label{Ek}
  \end{equation}   
where $m$ is the mass of the ball, $m_0$ is the mass of the origami, and the factor $\frac{1}{3}$ comes from the averaging. As for the potential energy, it can also be treated in terms of $x$ as a sum of the two energies (using the same notations):
  \begin{align}
  E_\mathrm{p}=E_{\mathrm{p,ball}}+E_{\mathrm{p,origami}}=m g x+m_0 g\frac{x}{2}.
  \label{Ep}
  \end{align}  

Having the expressions for the kinetic energy (\ref{Ek}), the potential energy (\ref{Ep}), and the effective torque (\ref{M(x)}), as well as the relation for the energy transformations in the system (\ref{En_trans}), we can consider the energy change per unit time that provides a new equation of motion similar to (\ref{eq_mot_one}):
  \begin{align}
  \left(m+\frac{m_0}{3}\right)\ddot{x}=\mathcal{M}(x,\dot{x})-\left(m+\frac{m_0}{2}\right)g,
  \label{eq_mot_fin}
  \end{align}
where $\mathcal{M}(x,\dot{x})$ is given by (\ref{M(x)}).

\section{Results}

\begin{figure}
\centering{\includegraphics[width=0.9\linewidth]{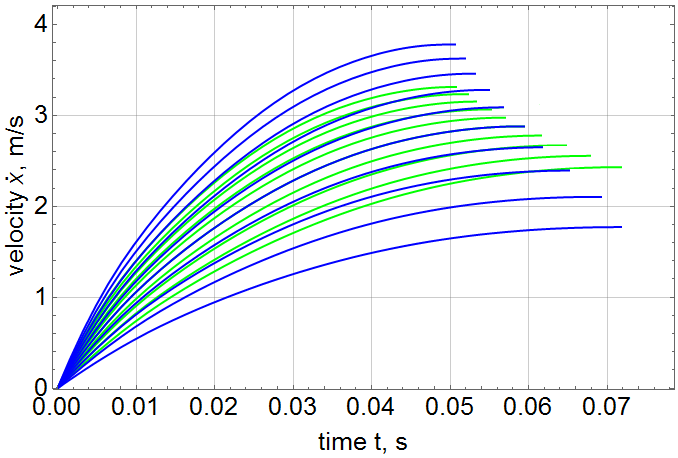}}
\vspace*{-0.2 cm}
\caption{Time-dependence of the vertical velocity $\dot{x}$ of the load obtained from the numerical solutions of the equation (\ref{eq_mot_fin}) for different numbers $N_1$ of horizontal and $N_2$ of vertical folds. Green lines show $N_1$ varying from 2 to 20 with fixed $N_2$=10, blue lines - vice versa. Naturally, when increasing the number of folds, the height of the ball also increases. The curves are cut off at the point of maximum, that is in the moment of separation from the construction.}
\label{solutions}
\end{figure}

The numerical solution of this equation gives the time dependence of the coordinate of the load and its vertical velocity (fig.~\ref{solutions}). It is influenced by all of the parameters of our system. As we recall, our task was to investigate the maximum height of the launched load, so we should find the velocity of the ball at the moment of its separation from the origami. Then we could easily convert it into the maximum potential energy and find the height of the ball. From our solutions, the maximal velocity of the ball can be obtained as a function of the governing parameters $N_1$, $N_2$, and the construction angle $\alpha$.

\subsection{Experiment}
\begin{figure}
\centering{\includegraphics[width=1\linewidth]{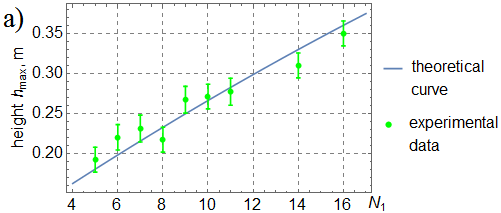}}
 \centering{\includegraphics[width=1\linewidth]{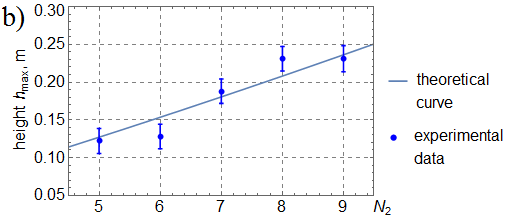}}
 \centering{\includegraphics[width=1\linewidth]{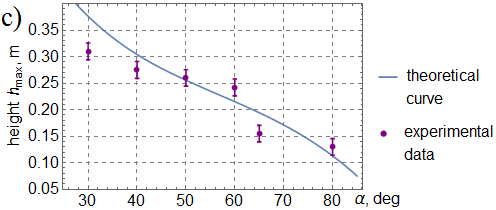}}
\vspace*{-0.6 cm}
\caption{Experimental dependences of the maximum height of the launched ball on the governing parameters: a) on the number $N_1$ of horizontal folds ($N_2$=7, $\alpha$=60$^{\circ}$); b) on the number $N_2$ of vertical folds ($N_1$=5, $\alpha$=60$^{\circ}$); c) on the construction angle $\alpha$ ($N_1$=7, $N_2$=7). Error bars in the plots are estimated approximately, to include the instrumental errors, the statistical distribution of repeated experiments, and the measurement inaccuracies related to the human factor.  Each experiment was repeated several dozen times, and then from the initial data set only vertical launches were selected. Each data set is compared to the theoretical solution with the corresponding parameters.}
\vspace*{-0.3 cm}
\label{exp}
\end{figure}

To study the dependence of the maximum height of the ball on the proposed parameters, we carried out a set of experiments varying independently $N_1$, $N_2$ and $\alpha$. We drew and printed three series of patterns with two parameters fixed and the third one varying in some range. All of the obtained dependences are shown in fig.~\ref{exp} and compared to the corresponding theoretical curves. As can be seen from these plots, the height of the launched load increases with increasing numbers of horizontal and vertical folds. This result is quite logical as the lever arm of the folds that decreases with increasing number of folds is not important for the resulting speed of the load. The energy coming from the creases is determined by their number and their inner properties. As concerns the impact of the construction angle (fig.~\ref{exp},~c), its increase results in a decrease of the height. Indeed, in this case the projection on the vertical axis of the force coming from the vertical creases becomes smaller and decreases the energy given to the load.

It should be noted that fixing some typical values of the parameters in the datasets does not allow us to reach the maximal height. To do this, we should perform the optimization while involving all of the parameters simultaneously. From our result, one can see that to increase the height of the ball it is necessary to increase the numbers of folds and to decrease the construction angle. Experimentally, we are naturally limited by our possibility to construct an origami with arbitrary parameters. When trying to reach the optimal parameters, the height approaches a value of the order of 50 cm. Our best result was achieved when we tried to make an origami corresponding to the extreme values of the parameters from our plots (fig.~\ref{exp}). Here, the load becomes unstable on top of the Miura-ori, so the vertical launch becomes complicated, and the full collection of statistics becomes cumbersome. Moreover, when the cells are so small, the impact of the human factor dramatically increases. The characteristic scale of construction errors becomes comparable to the size of the cells, so each inaccuracy results in significant changes to the launcher properties.

	\section{Discussion}
    It is impossible to disregard the fact that the standard flat Miura-ori studied in our paper is not the best construction for vertically launching a ball. We investigated it in order to understand the general principles of its work and to study the influence of different factors. 
    
    Our method was based on experimentally measured paper characteristics. That is why all of our experiments were carried out with the same batch of paper. When we tried to take another A4 paper, we got a strikingly different result: the same folding launches a ball almost two times higher. So, our theoretical functions cannot be applied to describe another type of paper. However, our treatment allows us to introduce arbitrary initial curves and parameters, and to perform all of the calculations again. Moreover, we can use the relations between the angles and the vertical coordinate to describe other Miura-like foldings. In general, the proposed method is applicable to the investigation of any kind of structure that can be treated as a sum of individual creases.
    
\begin{figure}
\center{\includegraphics[width=0.9\linewidth]{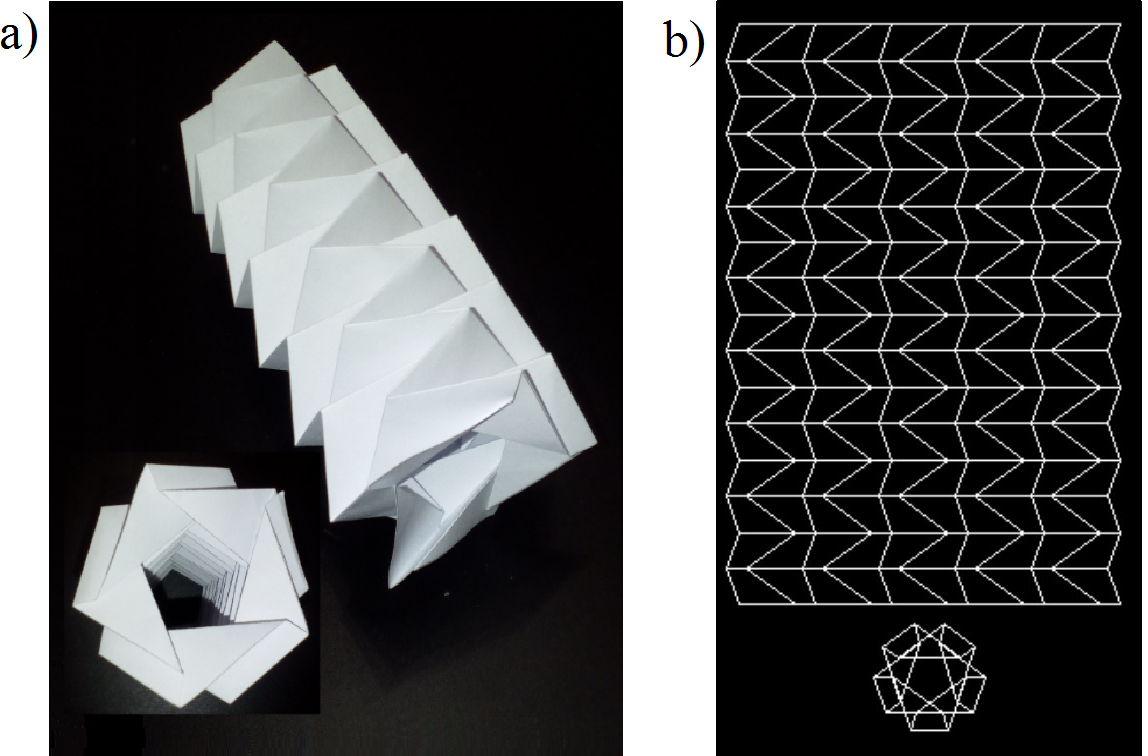}}
\vspace*{-0.1 cm}
\caption{The cylindrical Miura folding. a) Shape of the construction with the polygonal cylinder base. b) Scheme of this folding. One can notice that there are two different construction angles. These angles are related to each other and the number of edges of the polygonal base.}
\vspace*{-0.4 cm}
\label{fig.cyl}
\end{figure}

    For example, the simplest extension of a standard Miura-ori is the cylindrical Miura folding (fig.~\ref{fig.cyl}). The properties of such constructions have been studied in detail in \cite{Reid2017}. This type of pattern is particularly interesting for our problem because it provides a stable vertical launch of the ball and allows one to effectively use the energy stored in the folds. 
    
    The geometry of the cylindric folding is more complex than that of the flat one. In order to be cylindric, it already includes two different construction angles related with each other and a number of facets of the polygonal cylinder base. Thereby, the number of different angles between the pairs of faces and their derivatives with respect to the vertical coordinate increases. These relations become more complicated and imply some geometric restrictions caused by the ability of the folding to be easily compressed and decompressed. For example, for some values of angles, it is impossible to compress the cylinder by a continuous movement. It requires a kind of ``geometry breaking": to be compressed, the folding needs to be self-crossed, like the M\"{o}bius strip. As we are launching a ball, it is an important limitation in our problem, together with other one concerning breaking the  rigidity of some cylindrical constructions. That is, having two stable states, the origami needs its faces to be bent when getting from one stable state to another. These elastic modes, as well as other properties of the cylindrical Miura-ori are discussed in \cite{Reid2017}.
    
    This type of folding is able to provide a sufficiently large height of the ball (up to about 60 cm, which is slightly higher than our results for a flat Miura-ori) and can be equally well described by our theoretical model with the corresponding geometry.
    
\section{Conclusion}
	The elastic properties of the origami arise from the combination of its folds, so we first study the mechanics of an individual crease. Theoretically and experimentally, we determined the dynamics of a paper fold. In our experiments we noticed that paper has nonelastic and nonlinear properties. We saw the hysteresis loop on the static loading curve when opening and closing the fold. We experimentally measured the dependence of the angle, the angular velocity and the angular acceleration on time. Thereby, we found the torque provided by the paper fold and the dissipation in the fold depending on the angle and the angular velocity. Using this result, we solved the differential equation of motion and measured the dependence of the maximal angular velocity on the mass of the load.
    
    With the aid of the result for a single fold, we found the velocity of the entire origami, the velocity of the ball and the height reached by the ball. To do this, we created an algorithm to calculate the angles between the facets of a Miura-ori structure, and implemented it numerically.

	In addition, we have made a number of experimental measurements, in which we studied the velocity and the coordinate reached by the ball as a function of variable parameters, such as the number of horizontal and vertical folds, and the angle of the construction of the origami.
Comparing the experiment data to the numerical solution, we saw that the discrepancy was small. Still, if we use paper with other characteristics, its rigid properties will be different, and we will need to go through the fitting and calibration process once again.

	Our theoretical and experimental methods devised for the Miura-ori can be used to study other similar constructions.
	
	\section{Acknowledgements} We are very grateful to our friend Alexander Kostenko for revising the style of the manuscript, as well as to the referees and editors of the \textit{Emergent Scientist}, whose comments helped to substantially improve the quality of the article.

\section{Appendix 1} 
\subsection{Estimation of the energy loss of the load}
The load loses energy by interacting with air. With typical orders of magnitude of all the values in our problem, the estimated value of the Reynolds number of a load is
\begin{equation}
    \mathrm{Re}=\frac{\rho v D}{\eta}\sim \frac{1\frac{\mathrm{kg}}{\mathrm{m^3}} 1 \frac{\mathrm{m}}{\mathrm{s}}10^{-2} \mathrm{m}}{10^{-5}\frac{\mathrm{kg}}{\mathrm{m s}}}\sim10^3,
\end{equation}
so we will use the quadratic law to estimate the air drag force $F_{\mathrm{dr}}$. The energy dissipation is equal to the work of this force over the height of the launched load (the load is considered to be a ball)
\begin{equation}
     E_{\mathrm{dis}}\sim \overline{F_{\mathrm{dr}}}h
     \sim\frac{C \rho \pi r^2 \overline{v^2} h}{2}
     \sim \frac{C\rho \pi r^2 g h^2}{3}.
\end{equation}
The potential energy gain of the load is
\begin{equation}
     E_{\mathrm{p}}=m g h.
\end{equation}
The ratio of these two energies is (units of measurement are omitted, all the values are written out in the SI)
\begin{align}
   \nonumber \frac{E_{\mathrm{dis}}}{E_{\mathrm{p}}}&
   \sim \frac{C \rho \pi r^2 g h^2}{3 m g h}
   \sim \frac{C \rho \pi r^2 h}{3 m}\\
   &\sim \frac{0.5\cdot \pi \cdot 1\cdot 10^{-4}\cdot 0.5}{3\cdot 10^{-3}}\sim0.03
\end{align}

Thus, the energy dissipated in the air is two orders of magnitude less than the typical energies in our problem.
Finally, to estimate the energy of rotation of the load, we assume the load to be a ball making about one turn over the raising time, as was observed in our experiments.
\begin{align}
     \nonumber E_{\mathrm{rot}}&=\frac{I\omega^2}{2}
     \sim\frac{m r^2 (2 \pi/t)^2}{3}\\
     &\sim\frac{m r^2 (2 \pi)^2/(2h/g)}{3}
     \sim\frac{2\pi^2 m r^2 g}{3h},
\end{align}
so the ratio of this energy and the kinetic energy of the translational motion (which is almost equal to its potential energy gain) is (values are also given in SI)
\begin{align}
   \nonumber \frac{E_{\mathrm{rot}}}{E_{\mathrm{p}}}&\sim \frac{2\pi^2 m r^2 g}{3 m g h^2}=\frac{2\pi^2 r^2}{3h^2}\\
   &\sim\frac{2\pi^2\cdot 10^{-4}}{3\cdot0.5^2}\sim0.003.
\end{align}
Thus, the energy lost to the load rotation is also small.

\section{Appendix 2}
\subsection{Calculation of opening angles}
\begin{figure}
\centering{\includegraphics[width=1\linewidth]{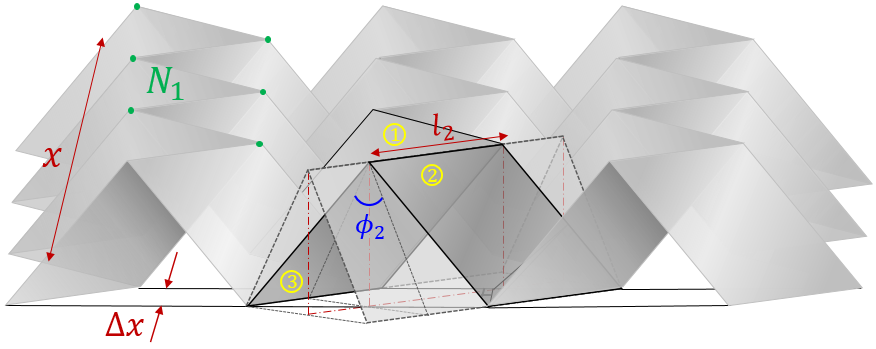}}
    \caption{A folded standard Miura sheet with the total displacement $x$ distributed over $N_1$ local displacements $\Delta x$. The opening angle $\phi_1$ is located for example between faces $1$ and $3$, and $\phi_2$ is respectively between faces $2$ and $3$ (the numbers of the faces are indicated in yellow circles).}
    \label{geom1}
\end{figure}

To find the opening angles as functions of the vertical coordinate $x$ let us consider a paper sheet with the length $L_1$ and the height $L_2$ folded using the standard Miura pattern with $N_1$ horizontal creases (fig.~\ref{geom1}). As the origami opens, the net displacement variable $x$ increases ($x\!=\!0$ for the completely closed origami). This total displacement is a sum of the local displacements $\Delta x$ of each cell, and we consider them to be equal ($\Delta x\!=\! x/N_1$). There are two different opening angles in this construction: $\phi_1$ near the ``horizontal" crease and $\phi_2$ near the ``vertical" crease.

\begin{figure}
\centering{\includegraphics[width=1\linewidth]{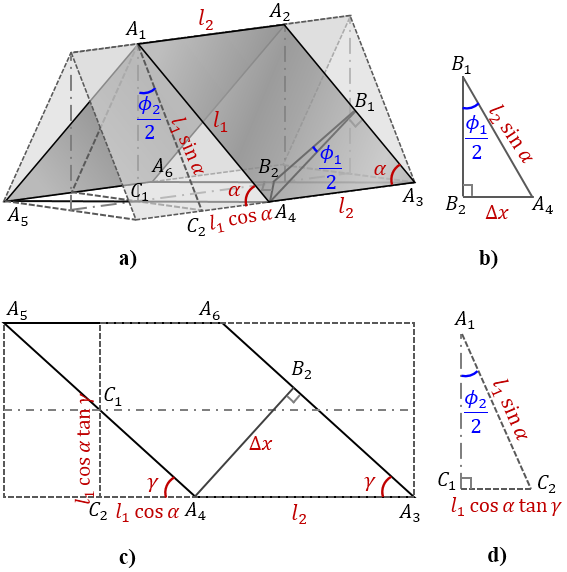}}
\vspace*{-0.7 cm}
    \caption{One Miura cell with the additional construction showing the two planes (a) and its separate elements (b, c, d). The planes $A_1 A_4 A_5$ and $A_2 A_3 A_6$ always remain parallel and in a real origami launcher are parallel to the ground. The local displacement $\Delta x$ is the distance between these planes.}
    \label{geom2}
\end{figure}

Let us focus on one particular cell (fig.~\ref{geom2}). It consists of parallelograms of sizes $l_2$ and $l_1$ (where $l_2\!=\!L_2/(N_1 \sin{\alpha})$) and has the construction angle $\alpha$.
From the right triangle $\triangle A_3 A_4 B_1$ (fig.~\ref{geom2},~a) one can find $A_4 B_1\!=\! l_2 \sin{\alpha}$. The line segment $A_4 B_2$ shows the distance between two parallel planes $A_1 A_4 A_5$ and $A_2 A_3 A_6$, so it is equal to the local displacement $\Delta x$. The angle between the planes $A_1 A_2 A_3$ and $A_2 A_3 A_6$ is equal to a half of the opening angle $\phi_1$ and can be found from the right triangle $\triangle A_4 B_1 B_2$ (see also fig.~\ref{geom2},~b). So,
\begin{equation}
    \sin{\frac{\phi_1}{2}}=\frac{\Delta x}{h \sin{\alpha}}=\frac{x}{L_2},
\end{equation}
which after a trigonometrical transformation results in eq.~(\ref{phi1}).

From the right triangle $\triangle A_1 A_4 C_2$ with \nolinebreak{$\angle A_1 A_4 C_2\!=\!\alpha$} we have \nolinebreak{$A_1 C_2\!=\!l \sin{\alpha}$}, and \nolinebreak{$A_4 C_2\!=\!l \cos{\alpha}$}. At the horizontal plane (fig.~\ref{geom2},~c) let \nolinebreak{$\angle A_4 A_3 B_2\!=\!\gamma$.} Then \nolinebreak{$\sin{\gamma}\!=\!\frac{\Delta x}{h}$.} As $A_4 A_5~||~A_3 A_6$, we get \nolinebreak{$\angle C_1 A_4 C_2\!=\!\gamma$}, so from the right triangle $\triangle C_1 A_4 C_2$ we obtain \nolinebreak{$C_1 C_2\!=\!l\cos{\alpha}\tan{\gamma}$.} Finally, in the right triangle $\triangle A_1 C_2 C_1$ (fig.~\ref{geom2},~d) the angle $\angle C_1 A_1 C_2$ is equal to one half of the opening angle $\phi_2$, so one can find

\begin{align}
    \nonumber\sin{\frac{\phi_2}{2}}&=\frac{l \cos{\alpha}\tan{\gamma}}{l\sin{\alpha}}=\frac{\cot{\alpha}\sin{\gamma}}{\sqrt{1-\sin^2\gamma}}\\
    &=\frac{\cos{\alpha}}{\sqrt{\frac{L_2^2}{x^2}-\sin^2\alpha}},
\end{align}
and for the doubled argument one can find eq.~(\ref{phi2}).

\end{document}